\documentclass[manuscript]{aastex}

\newcommand{\pr}{Phys. Rev.}
\newcommand{\npa}{Nucl. Phys. A}
\newcommand{\sci}{Science}

\newcommand{\cpl}{Chin. Phys. Lett.}
\newcommand{\aas}{Acta Astronom. Sin.}

\begin{document}

\title{Structures of the Vela pulsar and the glitch crisis from the Brueckner theory}

\author{A. Li\altaffilmark{1,2}, J. M. Dong\altaffilmark{2,3}, J. B. Wang\altaffilmark{4}, R. X. Xu\altaffilmark{5}}

\altaffiltext{1}{Department of Astronomy and Institute of Theoretical Physics and Astrophysics, Xiamen University, Xiamen, Fujian 361005, China; liang@xmu.edu.cn}
\altaffiltext{2}{State Key Laboratory of Theoretical Physics, Institute of Theoretical Physics, Chinese Academy of Sciences, Beijing 100190, China}
\altaffiltext{3}{Institute of Modern Physics, Chinese Academy of Sciences, Lanzhou 730000, China}
\altaffiltext{4}{Xinjiang Astronomical Observatory, Chinese Academy of Sciences, Urumqi, Xinjiang 830011, China}
\altaffiltext{5}{School of Physics and State Key Laboratory of Nuclear Physics
and Technology, Peking University, Beijing 100871, China}
\email{liang@xmu.edu.cn}

\begin{abstract}
Detailed structures of the Vela pulsar (PSR B0833-45, with a period of $89.33$ milliseconds) are predicted by adopting a recently-constructed unified treatment of all parts of neutron stars: the outer crust, the inner crust and the core based on modern microscopic Brueckner-Hartree-Fock calculations. To take the pulsar mass ranging from $1.0M_{\odot}$ to $2.0M_{\odot}$, we calculate the central density, the core/crust radii, the core/crust mass, the core/crustal thickness, the moment of inertia, and the crustal moment of inertia. Among them, the crustal moment of inertia could be effectively constrained from the accumulated glitch observations, which has been a great debate recently, known as ``glitch crisis''. Namely, superfluid neutrons contained in the inner crust, which are regarded as the origin of the glitch in the standard two-component model, could be largely entrained in the nuclei lattices, then there may not be enough superfluid neutrons ($\sim 4/5$ less than the previous value) to trigger the large glitches ($\Delta \nu/\nu_0 \sim 10^{-6}$) in the Vela pulsar. We then provide the first analysis of the crisis based on the microscopic basis, by confronting the glitch observations with the theoretical calculations for the crustal moment of inertia. We find that despite some recent opposition to the crisis argument, the glitch crisis is still present, which means that besides the crust superfluid neutrons, core neutrons might be necessary for explaining the large glitches of the Vela pulsar.
\end{abstract}


\section{Introduction}

The Vela pulsar (PSR B0833-45) is among the most studied pulsars, which are usually regarded as highly-magnetized, rapidly rotating neutron stars (NSs). By far there is an accurate determination of several pulsar properties, such as the distance~\citep{dis}, the spin period ($89.33$ milliseconds)~\citep{spin}, and 17 quasi-periodically produced glitches (sudden increases in spin frequency $\nu$)~\citep{Dodson02,Espinoza11,Yu13}. Those observations could be adequate for an accurate study of its global properties (mass, radius, moment of inertia) and inner structures.
On the other hand, thanks to the quick development of modern nuclear many-body theories, we are now in a position to describe properly the overall NS structures from its outer crust to the inner core, with a density range from less than $ 10^{-10} $ fm$^{-3}$ to larger than 1 fm$^{-3}$, to be compared with the the nuclear saturation density of 0.17 fm$^{-3}$.

The present study aims to provide the first detailed theoretical structure study of the Vela pulsar. The calculations are based on a recently-constructed Barcelona-Catania-Paris-Madrid (BCPM) nuclear energy density functional~\citep{bcpm} from modern microscopic Brueckner-Hartree-Fock (BHF) calculationsl~\citep{book}. It is the first equation of state (EoS) for NSs from the outer crust to the core based on modern microscopic calculations. It can accord with the ground-state properties of nuclei along the periodic table, the heavy-ion flow investigations~\citep{flow}, and the observational constrains from the 2 recent precisely-measured heavy pulsars¡¯ masses~\citep{2mass10,2mass13}, which could be seen as the the most well-accepted three constrains nowadays in the market. A brief introduction of the BHF model will be presented later. The adapted nuclear interaction is Argonne v18 potential plus nucleonic three-body forces computed with the Urbana model.

Actually, other unified EoSs for NSs appeared also based on a phenomenological basis, such as the Douchin $\&$ Haensel model~\citep{dh01} and the BSk family derived by the Brussels-Montreal group~\citep{chamel11,pearson12,fantina13,potekhin13}. They could also describe sufficiently well the experimental data mention above. The most recent BSk21 and BSk20 Skyrme nuclear effective forces could yield the 2 recent precisely-measured heavy pulsars¡¯ masses~\citep{2mass10,2mass13} and are chosen as a representative example of contemporary EoS for the complete NS structure and a comparison with the BCPM results. We mention here that the high-density part of the BSk21 EoS was adjusted to the result of the BHF calculations~\citep{LS08} using also Argonne v18 potential but plus a microscopic nucleonic three-body forces, and the high-density part of the BSk20 EoS was fitted to the Akmal-Pandharipande-Ravenhall EoS~\citep{apr} which has been widely used and serves as a baseline EoS
in some recent articles (e.g., Kurkela et al. 2010).

Also, before the success of a unified NS EoS, one usually combined two or three EoSs which handle only some density region of the star, respectively. For example, one could use the BHF calculations to obtain the EoS of the high-density core, and match it with the ones by Negele and Vautherin (NV)~\citep{nv} in the medium-density regime  ($0.001~$fm$^{-3}<\rho<0.08~$fm$^{-3}$), and the ones by Feynman-Metropolis-Teller~\citep{fmt} and Baym-Pethick-Sutherland (BPS)~\citep{bps} for the outer crust ($\rho<0.001$~fm$^{-3}$). It could be also interesting to examine the accuracy of this kind of combined EoS (labelled as Av18* in the following). Therefore totally four cases of the EoSs are used in the present study, and only for the first three we provide detailed calculated values as predicted properties for the Vela pulsar.

The second problem we would like to address here is the ``glitch crisis'' related to the moment of inertia of the stellar crust. We will explain more clearly this issue in the following three paragraphs.

``Glitch crisis'' was first proposed independently by Andersson et al. (2012) and Chamel (2013). It says that due to a large effective neutron mass (so-called \emph{crustal entrainment} effect by the crystal lattice) in the superfluid,  then glitches observed in the Vela pulsar can not be explained in the standard two-component models.

Specifically, in the standard two-component mechanism~\citep{Baym69}, the superfluid neutrons are limited in the inner crust of the star, where the density has exceeded the neutron drip density. It is a non-uniform system composed of neutron-rich nuclei, electrons and superfluid neutrons. Those neutron superfluid vortices are decoupled from the rotational evolution of the charged component of the crust (and that part of the core that couples strongly to it), since their interactions with the lattice of nuclei make them pinned to the crust. The superfluid component and the charged component will recouple when the lag between the two angular frequencies reaches a critical value, accompanied by a transfer of angular momentum from the superfluid component to the charged component. Then a glitch happens.

From the observed glitch activity of the Vela pulsar, one can derive a minimum fractional moment of inertia of the angular momentum reservoir $\Delta I$ compared to that of the
portion of the star it couples to at the time of glitch $I$, $\Delta I/I \geq 1.6 \%$. In the standard two-component model, this leads to $I_{\rm c}/I \geq 1.6 \%$,
where the subscript $c$ referring to the superfluid component from the crust part. The superfluid can be entrained by the crustal lattice, which severely limits the maximum number of superfluid neutrons that might transfer angular momentum during glitches. The ratio of free neutrons to total superfluid neutrons was estimated to be around $0.2$~\citep{crisis12,crisis13}, then the glitch constrain is updated to be $I_{\rm c}/I \geq 7 \%$. With this new constrain of the fractional moment of inertia, the pulsar could not be heavier than $\sim1.0M_{\odot}$~\citep{crisis12,crisis13}, leaving a doubtful situation that whether they are still likely to be formed in a type II supernova explosion as generally believed. This is known as ``glitch crisis''.

Several illuminating works~\citep[e.g.,][]{Guger14,Link14,Piekarewicz14,Hooker15,Li15a,Li15b} are followed and have contributed a more comprehensive understanding of this problem. The argument of ``glitch crisis'' relies on the assumption that the entire core of the star is coupled to the crust when glitches happen. Some studies pointed out that the glitch rise timescale ($<40$s~\citep{Dodson02}) might be much smaller than the core-crust coupling time (about the order of days), which means only a fraction (maybe a half or less) of the core part could be coupled to the charged crust part and contribute to the total stellar moment of inertia~\citep{Link14}. If it is the case, there would be no crisis.
However, a detailed determining of the core-crust coupling timescale included at least the type of proton superconductivity in the core~\citep{Link14}, the details of how the vortices react to the pinning force, the realistic modeling of the pinning force in the crust~\citep{Pizzochero11,Haskell13}, and the impact of nuclear ``pasta'' phases near the crust bottom~\citep{crisis13}, which are presently very uncertain. An effort to microscopically determine the pinning force are undergoing~\citep{Li15}.

Some studies~\citep[e.g.,][]{Piekarewicz14,Li15a} argued that the uncertainties in the EoS of neutron-rich matter are large enough to accommodate theoretical models that predict large crustal moments of inertia, and if there is large enough transition pressures, large Vela glitches could be generated without invoking an additional angular-momentum reservoir beyond that confined to the solid crust. Their calculations are based on three combined EoSs corresponding to the outer crust, the inner crust and the liquid core, respectively. Namely, a BPS component mentioned above is used for the outer crust up to the neutron drip density, and the crust-core transition density and the EoS of the liquid core are obtained from the phenomenological relativistic mean-field (RMF) models , and a polytropic interpolation is used for the the inner crust between the neutron drip density and the crust-core transition density.

Since we have in hand a consistent determination of not only the crust EoS and the core EoS, but also the crust-core transition properties from one microscopic basis, we would like to revisit carefully the crisis argument, to see whether the crustal superfluid is enough or not. Because our adapted model has a much reduced
set of model parameters~\citep{bcpm}, and there is no sensitivity on the density dependence of the symmetry energy and the crust-core transition density, the result may have important implications for both NS physics and pulsar observations. To serve as the present purpose, here we omit the possible effect of pasta phase mentioned above, as well as that of the strangeness phase transitions in the star's core~\citep[e.g.,][]{Li04k,Li06k,Li08q,Li10k,Li11y,Li14y,Li15q}. Detailed studies on them could be also very interesting and shall be left for future studies.

The paper is organized as follows. In section 2, we introduce briefly the adapted NS models; In section 3, numerical results are discussed, including various structural properties of the star, and an analysis of the crustal moment of inertia, followed by several tabular data of the star structure predictions. A short summary is finally presented in section 4.

\section{The model}

As usual, we describe a NS as a spherically symmetric object in hydrostatic equilibrium, with gravity balanced by the pressure produced by the compressed nuclear matter. From the surface to its core, with the increase of density, the following regions appear as follows:\\
$1)~Outer~crust$ : Non-uniform coulomb lattice of neutron-rich nuclei embedded in a degenerate electron gas;\\
$2)~Inner~crust$ : Non-uniform system of more exotic neutron-rich nuclei, degenerate electrons and superfluid neutrons;\\
$3)~Core$ : Uniform nuclear matter in weak-interaction equilibrium (or $\beta-$equilibrium) with leptons.

The BPCM energy density functional~\citep{bcpm} can handle the EoSs of both the core and the crust regions in the same microscopic physical framework, and serve as an adequate method of the present purpose. For completeness, we briefly describe in the following how the three parts of NSs are modelled in the BPCM model, which originates from the microscopic BHF theory.

The BHF theory~\citep{book} is currently one of the most advanced microscopic approaches to the EoS of nuclear matter. It uses realistic free nucleon-nucleon (NN) interaction as input, with parameters fitted to NN scattering phase shifts in different partial wave channels and to properties of the deuteron. The inclusion of the microscopic TBF in the calculation for dense stellar matter and NS properties have been performed in many works~\citep[e.g.,][]{book,Li04k,Li06k,Li08q,Li10k,Li11y,Li14y,Li15q}.

The BHF calculations can provide microscopically the EoS of the high-density NS core part, but the crust EoS is currently out of its reach since the NS crust includes various inhomogeneous structures and the electromagnetic force should also be included. In order to connect directly the crust calculations to the microscopic BHF results for homogeneous nuclear matter, the BPCM energy density functional is built up with a bulk part obtained from the BHF calculations for symmetric and neutron matter, added in usual ways by the phenomenological surface part, the Coulomb part, the spin-orbit part, and the pairing contributions. Then the ground-state properties of nuclei in the outer crust are calculated in the Wigner-Seitz (WS) approximation and reproduced if they are known experimentally. The inner crust is computed using self-consistent Thomas-Fermi calculations with the BCPM functional in different WS configurations, where the low-density neutron
gas and the bulk matter of the high-density nuclear structures
are given by the same microscopic BHF calculation. Finally the NS core is assumed to be composed of normal ($npe\mu$) matter and is calculated from the BHF calculations together with the $\beta-$equilibrium condition and the charge-neutrality constraint. One may refer to Li et al. (2015) for a detailed demonstration of this procedure.

In Fig.~1, four cases of the NS EoSs (BCPM, BSk21, BSk20, Av18*) adapted in the present work are shown. Namely the total pressures (in the units of MeV/fm$^3$) are plotted as a function of the number density (in the units of fm$^{-3}$). One could use other units after making necessary conversions, for example, 1 erg/cm$^3$ = 1.6022$\times 10^{33}$ MeV/fm$^3$, 1 cm$^{-3}$ = $10^{39}$ fm$^{-3}$. In the left (right) panel of the figure a large (small) range of density is displayed. It is clearly seen in the left panel that the BCPM (Av18*) EoS is the softest (stiffest) among the four, and also the BSk21 EoS is stiffer than the BSk20 one. The differences around the crust-core transitions for the fours cases are displayed in the right panel, with four dots marking the corresponding threshold densities. The matched Av18* EoS has some nonsmooth behaviours near the crust-core threshold.

We collected also the numbers of the crust-core transition properties (such as the transition density and the transition pressure) in Table 1. The four transition pressures are all larger than the largest case (0.692 MeV fm$^{-3}$) presented in the RMF calculations~\citep{Piekarewicz14} mentioned in the introduction, and can be regarded as large enough transition pressures. However, as will shown later, large enough transition pressures could not provide large crustal moments of inertia, hence the glitch crisis may not be avoided after all.

The last part of this section is devoted to the equations of the stellar structure and its moments of inertia.

The stable configurations of a non-rotating NS can be obtained from the well known spherical-symmetry imposed Tolman-Oppenheimer-Volkoff equations~\citep{tov} for the pressure $P$ and the enclosed mass $M$
\begin{equation}
 \frac{dP(r)}{dr}=-\frac{GM(r){\Large{\varepsilon }}(r)}{r^{2}}
 \frac{\Big[1+\frac{P(r)}{{\Large{ \varepsilon }}(r)}\Big]
 \Big[1+\frac{4\pi r^{3}P(r)}{M(r)}\Big]}
 {1-\frac{2GM(r)}{r}},
    \label{tov1:eps}
\end{equation}
\begin{equation}
\frac{dM(r)}{dr}=4\pi r^{2}{\Large{\varepsilon }}(r),
    \label{tov2:eps}
\end{equation}
once the EoS $P({\Large{ \varepsilon }})$ is specified, being ${\Large{ \varepsilon }}$ the
total energy density ($G$ is the gravitational constant). For a
chosen central value of the energy density, the numerical
integration of Eqs. (\ref{tov1:eps}, \ref{tov2:eps}) provides the
mass and radius of the star.

Assuming the star is rotating uniformly with a stellar frequency $\Omega$ far smaller than the Kepler frequency at the equator ($\Omega << \Omega_{max} \approx \sqrt{GM/R^3}$), the moment of inertia of a star with a radius $R$ and an angular frequency $\Omega$ can be calculated in the slow-rotation approximation based on the above spherical-symmetry metric combined with an axis-symmetry perturbation~\citep{slow}:
\begin{equation} \label{eq:MoI1}
I=\frac{8 \pi}{3} \int_0^R r^{4}e^{-\nu(r)}\frac{\bar{\omega}(r)}{\Omega}\frac{\left( \mbox{$ \varepsilon $} (r)+P(r) \right)}{\sqrt{1-2GM(r)/r}}\mathrm{d}r,
\end{equation}
where $\nu(r)$ is a radially-dependent metric function given by
\begin{equation} \label{eq:MoI2}
\nu(r)=\frac{1}{2} \ln \left( 1- \frac{2GM}{R} \right) - G \int_r^R \frac {\left( M(x)+4 \pi x^{3} P(x) \right)}{x^{2} \left( 1 - 2 GM(x)/x \right)} \mathrm{d}x,
\end{equation}
and $\bar{\omega}$ is the frame dragging angular velocity given by
\begin{equation} \label{eq:MoI3}
\frac{1}{r^{3}}\frac{\mathrm{d}}{\mathrm{d}r} \left( r^{4}j(r)\frac{\mathrm{d}\bar{\omega}(r)}{\mathrm{d}r} \right) + 4 \frac{\mathrm{d}j(r)}{\mathrm{d}r} \bar{\omega}(r)=0,
\end{equation}
where
\begin{equation} \label{eq:fp3}
j(r)=e^{-\nu(r)-\lambda(r)}=\sqrt{1-2GM(r)/r} e^{-\nu(r)},
\end{equation}
for $r \leq R$.

Since we follow here the standard two-component model, the fractional moment of inertia responsible for a glitch is the crustal moment of inertia from the inner crust of the star, which is given by
\begin{equation} \label{eq:new1}
I_{\rm c}=\frac{8 \pi}{3} \int_{R_{\rm c}}^{R} r^{4}e^{-\nu(r)}\frac{\bar{\omega}(r)}{\Omega}\frac{\left( \mbox{$ \varepsilon $} (r)+P(r) \right)}{\sqrt{1-2GM(r)/r}}\mathrm{d}r
\end{equation}
where $R_{\rm c}$ is the crust-core transition radius.

The mass-radius relations of the star are shown in the left panel of Fig.~2, with the star mass as a function of the central density shown in the right panel, four cases of the NS EoSs (BCPM, BSk21, BSk20, Av18*) adapted in the present work. All the four EoSs could support NSs as heavy as $2M_{\odot}$, fulfilling the constraints from the 2 recent precisely-measured heavy pulsars¡¯ masses~\citep{2mass10,2mass13}. Also, we obtain the smallest maximum mass in the BCPM case as expected from the underlying soft EoS (shown in Fig.~1).

\section{Results}
We are now ready to study in specific the detailed structures of the Vela pulsar. It has a measured period of $89.33$ milliseconds. The mass is unknown, but could be reasonable concluded to be between $1.0M_{\odot}$ and $2.0M_{\odot}$. We then calculate various properties (the central density, the core/crust radii, the core/crust mass, the core/crustal thickness, the moment of inertia, and the crustal moment of inertia) as a function of the mass ranging from $1.0M_{\odot}$ to $2.0M_{\odot}$.

From Fig.~2, we can first conclude that its radius should be around $11 - 12$ km based on the BCPM EoS. For a $1.4M_{\odot}$ star, it is 11.75 km, comfortably lying between the narrow range of $10.7 - 13.1$ km based on an elaborate combined analysis \citep{lattimer13} from nuclear masses, neutron skins, heavy ion collisions, giant dipole resonances, and dipole polarizabilities in laboratory experiments. The central number density of 0.548 fm$^{-3}$, about 3 times of the nuclear saturation density (0.17 fm$^{-3}$).

In the left panel of Fig.~3, the crustal mass and the core mass are shown separately as a function of the stellar mass for two representative cases of NS EoSs (BCPM, BSk21). For both the BCPM EoS and the BSk21 EoS presented here, with the increase of the stellar mass, its inner density becomes high, hence the mass contribution from the core part is enhanced, even through the star itself should get small, as seen from the right panel of the same figure, where the stellar radius and the crustal thickness is plotted also as a function of the stellar mass; Also, heavier stars may have thinner crusts, leading to a smaller crustal mass. We have checked that for other EoSs, the conclusions are the same.

Actually, the inner crust thickness ($R_{icrust}$), defined as
the distance between the neutron drip density and the core-crust
transition density, is one of the crucial quantities for deriving the density-dependent behaviour of the nuclear symmetry energy, which is both controversial and important to fundamental physics and astrophysical observations~\citep[e.g.,][]{lattimer13}. Since the inner crust thickness depends primarily on the stellar radius $R$, $R_{icrust}/R$ could be more meaningful. We then plot $R_{icrust}$ and $R_{icrust}/R$ in Fig.~4 as a function of the stellar mass for all the four cases of NS EoSs (BCPM, BSk21, BSk20, Av18*) adapted in the present work. We see that, like the stellar radius $R$, the inner crust thickness $R_{icrust}$ decreases with the stellar mass, and the ratio $R_{icrust}/R$  is also a decreasing function with the stellar mass. The general decreasing behaviour of both the four EoSs are very similar. Quantitatively, we find that with the three unified EoSs (BCPM, BSk21, BSk20), $R_{icrust}$ ($R_{icrust}/R$) lies in the range of $0.2 - 0.84$ km ($0.07 - 0.02$), and for a $1.4M_{\odot}$ NS, $R_{icrust} = 0.53$ km, $R_{icrust}/R = 0.045$. There values are lower than the predictions made by Lattimer \& Lim (2013), where with 1$\sigma$ errors, they get $R_{icrust} = 0.96^{(+0.17)}_{(-0.13)}$ km and $R_{icrust}/R = 0.082\pm0.08$.

Next, in Fig.~5 the results of the moments of inertia are plotted as a function of the stellar mass for both the total ones $I$ and the crustal ones $I_{\rm c}$, with all the four cases of NS EoSs (BCPM, BSk21, BSk20, Av18*) adapted in the present work. The total (crustal) moments of inertia is with an order of magnitude of $10^{45}$ g cm$^2$ ($10^{43}$ g cm$^2$ ). We see a general feature that the total moments of inertia increase with the stellar mass, while the crustal moments of inertia do the opposite. Among the four EoSs presented here, the BCPM one shows a slightly rapider change with the stellar mass than the other three.

From Fig.~5, we could except that the fractional moments of inertia $I_{\rm c}/I$ should be a decreasing function with the stellar mass. This is demonstrated in Fig.~6, with all the four cases of NS EoSs (BCPM, BSk21, BSk20, Av18*) adapted in the present work. We first notice that for the previous glitch constraint $I_{\rm c}/I \geq 1.6\%$~\citep{Link99}, the mass of the Vela pulsar should only be smaller than 1.8M$_{\odot}$ based on selected NS EoS models here. This could be easily fulfilled for an isolated pulsar~\citep{Lattimer2007}. However, the discovered crustal entrainment shifts the constrain to around $I_{\rm c}/I \geq 7\%$~\citep{crisis12,crisis13}, which could hardly be fulfilled by a pulsar heavier than $1.0 M_{\odot}$ as seen from Fig.~6. Actually the predicted fractional moments of inertia based on three unified NS EoSs (BCPM, BSk21, BSk20) are no more than $6.6 \%$, therefore we conclude that the standard two-component model, namely only crustal superfluid neutrons are responsible for glitches, could not explain the large glitches quasi-periodically produced in the Vela pulsar. One might also consider the core neutrons as necessary.

Finally at the end of this section, taking the pulsar mass ranging from $1.0M_{\odot}$ to $2.0M_{\odot}$, various structure properties (the central density, the masses, radii, the moments of inertia of the core part, the inner-crust part, and the outer crust part) are tabulated in details in Tables 2-4, for the BCPM EoS, the BSk21 EoS, and the BSk20 EoS, respectively.

\section{Summary}
In this paper, we have made use of the contemporary EoSs for the complete NS structure, especially the recently-derived BCPM one based on modern microscopic BHF calculations, and have predicted the detailed structures of one of the most important isolated pulsars, the Vela pulsar (PSR B0833-45) (with a period of $89.33$ milliseconds). For this purpose, we adapt the the slow-rotation approximation, and solve the equations of the stellar structure for its mass, radius and moment of inertia from the underlying chosen EoS. Three unified NS EoSs (BCPM, BSk21, BSk20) are employed in the calculations, to be compared with a matched EoS of Av18* often used in the literature.

We find that for a typical mass of $1.4 M_{\odot}$, the star has a central density of $0.548$ fm$^{-3}$, with the radius $R$ of 11.75 km and the inner crustal thickness $R_{icrust}/R = 0.045$, which mostly agree with the previous results. Various structure properties are tabulated for three contemporary unified NS EoSs (BCPM, BSk21, BSk20), including the central density, the masses, radii, the moments of inertia of the core part, the inner-crust part, and the outer crust part.

Furthermore, based on the calculated results of the moments of inertia, we provide the first analysis of the ``glitch crisis'' problem from a microscopic unified NS EoS. We find that the resulting fractional moments of inertia from the crust part could be no more than $6.6 \%$, which contradicts with the conclusion that this value should be larger than $7 \%$ according to the glitch observations of the Vela pulsar. Therefore, we argue that the crisis are still present in the standard two-component model, opposite with other recent studies.

\begin{acknowledgements}
We would like to thank Dr. G. F. Burgio and Dr. H.-J. Schulze for valuable discussions; One of us (AL) would also like to thank their hospitality during her stay in Catania. The work was supported by the National Natural Science Foundation of China
(Nos. 11225314, 11403086, 11405223, U1431107) and the West Light Foundation of the Chinese Academy of Sciences (CAS) (No. XBBS201322).
\end{acknowledgements}

\begin{figure}
\epsscale{1.0}
\plotone{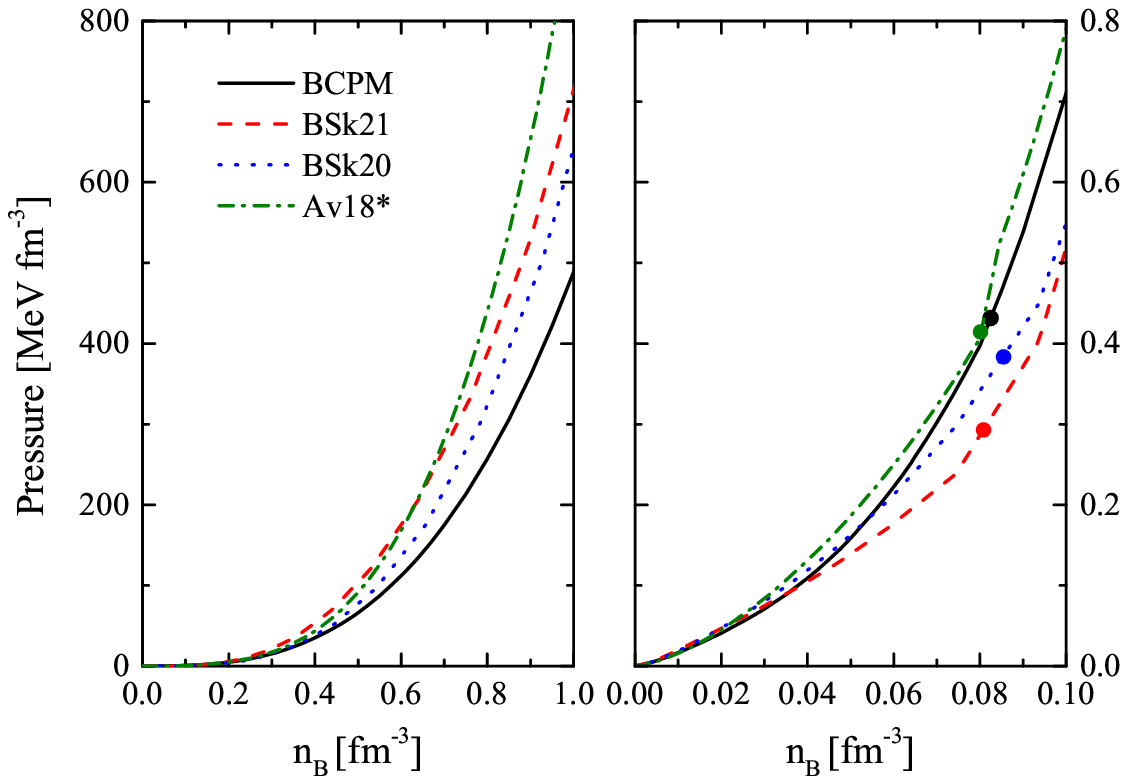}
\caption{(Color online) Pressure as a function of the number density for the four cases of NS EoSs (BCPM, BSk21, BSk20, Av18*) adapted in the present work, with a large (small) range of density displayed in the left (right) panel. Four dots in the right panel mark the crust-core threshold densities in four cases. See text for details.}\label{fig1}
\end{figure}

\clearpage

\begin{figure}
\epsscale{1.0}
\plotone{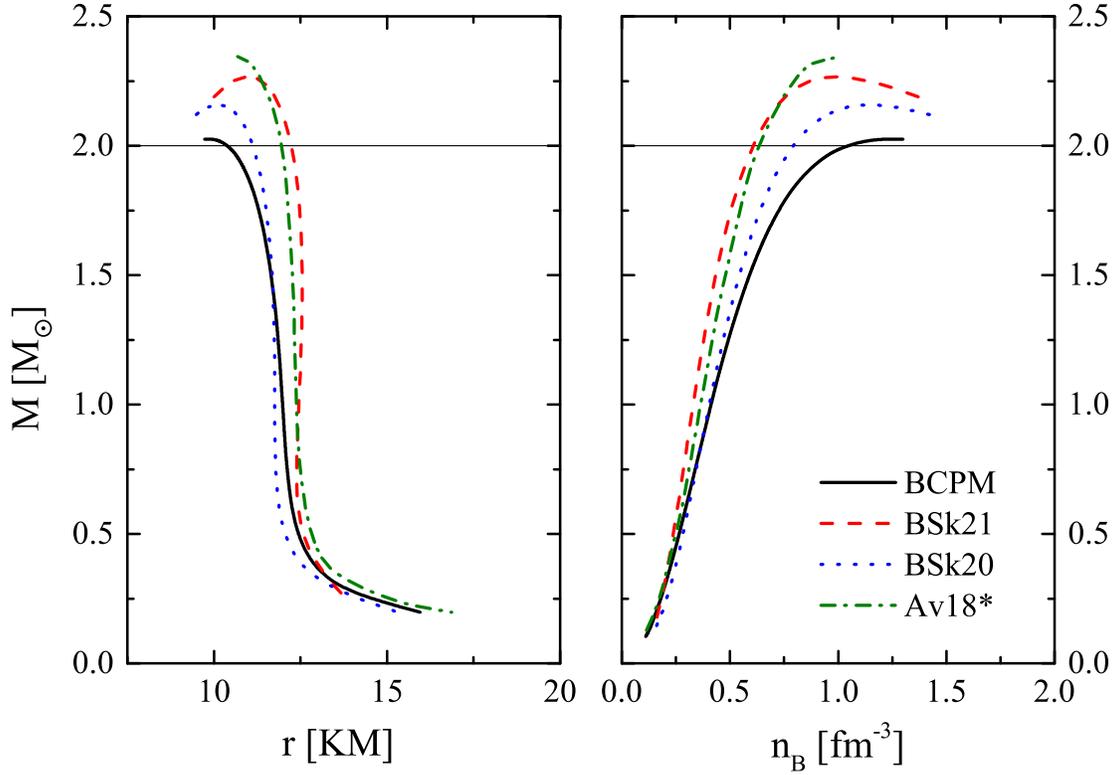}
\caption{(Color online) Mass-radius relations of the star (left panel) and the star mass as a function of the central density (right panel), for four cases of NS EoSs (BCPM, BSk21, BSk20, Av18*) adapted in the present work. The horizontal line is the recent 2-solar-mass constraint from the mass measurements of PSR J1614-2230 and PSR J0348+0432~\citep{2mass13,2mass10}.}\label{fig2}
\end{figure}

\clearpage

\begin{figure}
\epsscale{1.0}
\plotone{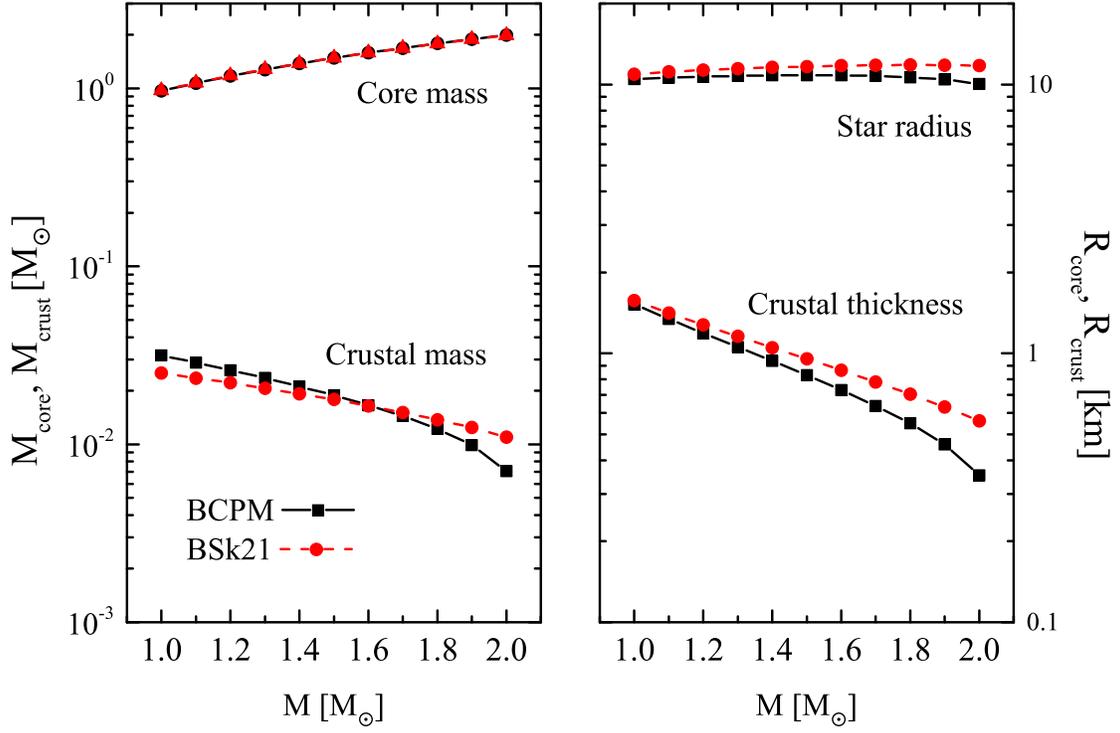}
\caption{(Color online) (Left panel) NS crustal mass and the core mass as a function of the stellar mass for two cases of NS EoSs (BCPM, BSk21); (Right panel) Same with the left panel, but with the crustal thickness and the stellar radius.}\label{fig3}
\end{figure}

\clearpage

\begin{figure}
\epsscale{1.0}
\plotone{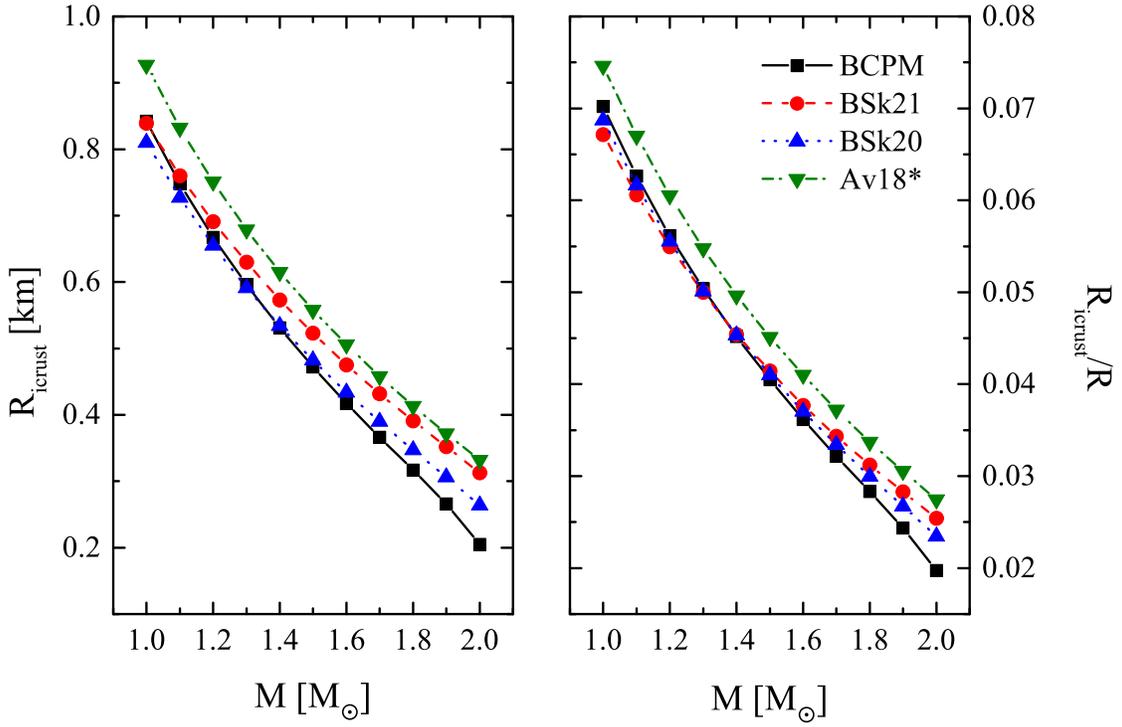}
\caption{(Color online) Inner crust thickness $R_{icrust}$ (left panel) and $R_{icrust}/R$ (right panel) as a function of the stellar mass for all the four cases of NS EoSs (BCPM, BSk21, BSk20, Av18*) adapted in the present work.}\label{fig4}
\end{figure}

\clearpage

\begin{figure}
\epsscale{1.0}
\plotone{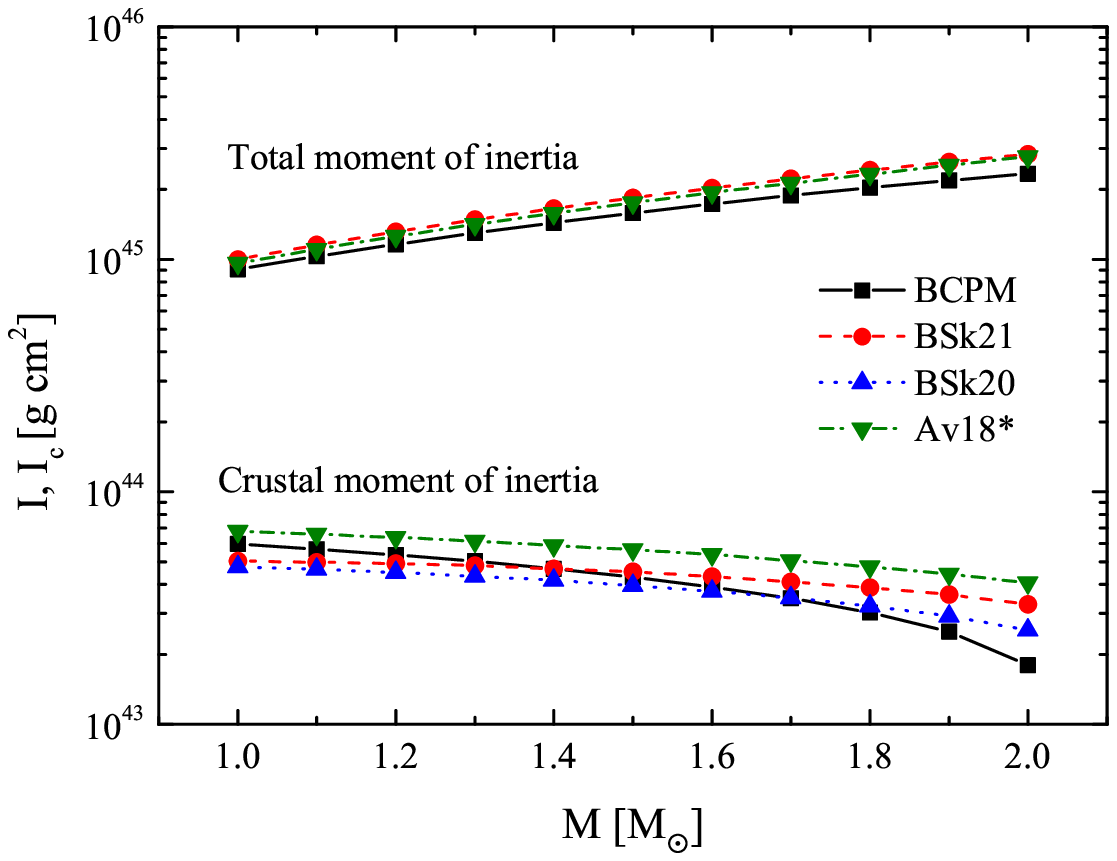}
\caption{(Color online) Moments of inertia as a function of the stellar mass for both the total ones $I$ and the crustal ones $I_{\rm c}$, with all the four cases of NS EoSs (BCPM, BSk21, BSk20, Av18*) adapted in the present work.}\label{fig5}
\end{figure}

\clearpage

\begin{figure}
\epsscale{1.0}
\plotone{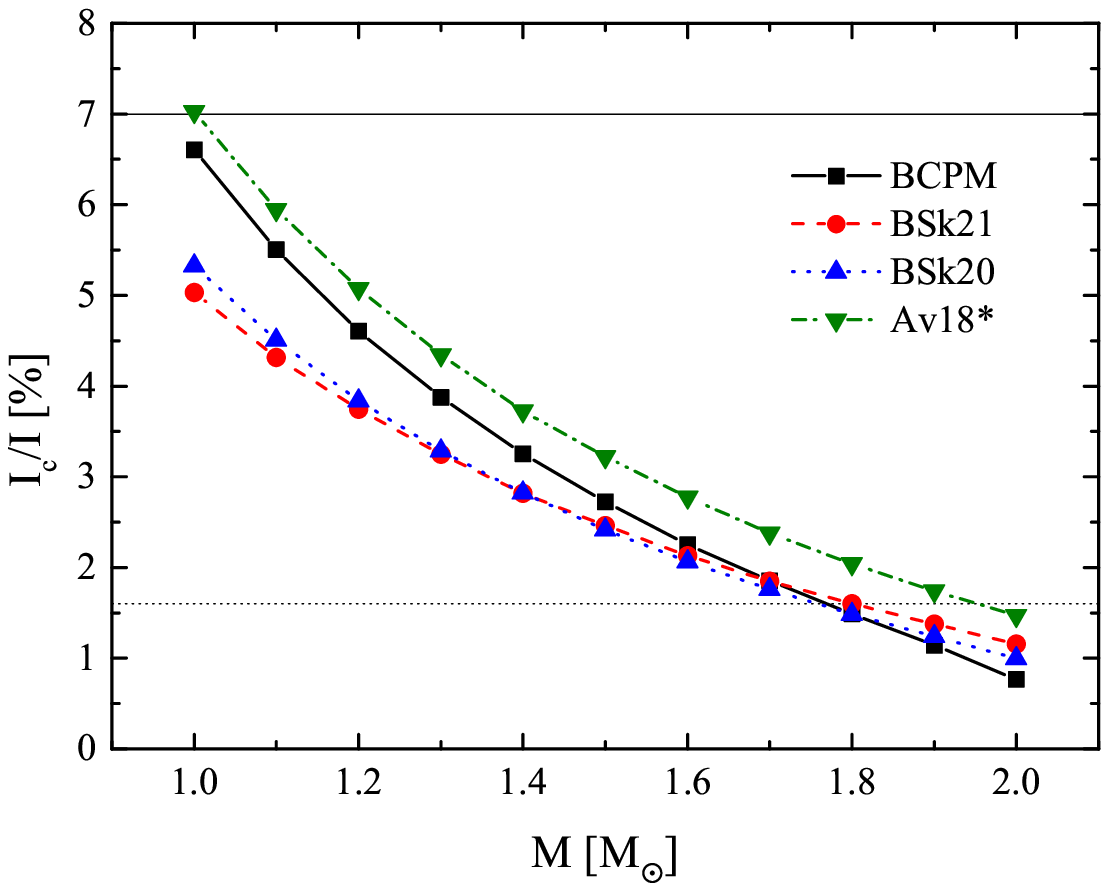}
\caption{(Color online) Fractional moments of inertia as a function of the stellar mass for both the total ones $I$ and the crustal ones $I_{\rm c}$, with all the four cases of NS EoSs (BCPM, BSk21, BSk20, Av18*) adapted in the present work.}\label{fig6}
\end{figure}

\clearpage

\begin{table}
\tabcolsep 0pt
\caption{Crust-core transition properties (the transition density $n_{c}$, the transition pressure $P_{c}$) for four cases of NS EoSs (BCPM, BSk21, BSk20, Av18*) adapted in the present work.}
\vspace*{-12pt}
\begin{center}
\def\temptablewidth{0.5\textwidth}
{\rule{\temptablewidth}{1pt}}
\begin{tabular*}{\temptablewidth}{@{\extracolsep{\fill}}ccc}
Model & $n_{c} $ [fm$^{-3}]$ &$P_{c} $ [MeV fm$^{-3}]$ \\   \hline
BCPM & 0.0825  &  0.4317  \\
Bsk21 & 0.0809 &   0.2952   \\
Bsk20 & 0.0854 &  0.3839     \\
Av18* & 0.0800  & 0.4059
       \end{tabular*}
       {\rule{\temptablewidth}{1pt}}
       \end{center}
       \end{table}

\clearpage

\begin{table}[!htb]
\begin{center}
\caption{Predictions for the properties of the Vela pulsar (PSR B0833-45) with a spin period of 89.33 milliseconds~\citep{spin}, taking the pulsar mass ranging from $1.0M_{\odot}$ to $2.0M_{\odot}$. The cental densities are in unites of fm$^{-3}$, the masses in solar masses $M_{\odot}$ (except the outer crust mass $M_{\rm ocrust}$ in $10^{-5}M_{\odot}$), the radii in kilometers, and the total moments of inertia in $10^{45}$g cm$^2$. Calculations are done with the BCPM NS EoS~\citep{bcpm}, which is the first unified NS EoS from the outer crust to the core based on a microscopic basis (see text for details).}\label{bcpm}
\begin{tabular}{cc ccc cccc cc}
\hline\hline
            {Mass~}   {Cent.}
            &\multicolumn{3}{c}{Mass }
            &\multicolumn{4}{c}{Radius}
            &\multicolumn{2}{c}{Moment of inertia} \\
            &{core}&{icrust} &{ocrust}
            &{total}&{core} &{icrust} &{ocrust}
            &{total}&{fraction} \\
\hline
             {1.0~~} {0.412}
            &{0.97}&{0.032} &{4.91}
            &{12.00}&{10.47} &{0.84} &{0.68}
            &{0.905} &{0.066}   \\
             {1.1~~} {0.443}
            &{1.07}&{0.029} &{4.37}
            &{11.95}&{10.60} &{0.75} &{0.59}
            &{1.031} &{0.055}   \\
             {1.2~~} {0.476}
            &{1.17}&{0.026} &{3.85}
            &{11.89}&{10.70} &{0.67} &{0.52}
            &{1.162} &{0.046}   \\
             {1.3~~} {0.511}
            &{1.28}&{0.024} &{3.39}
            &{11.83}&{10.77} &{0.60} &{0.46}
            &{1.297} &{0.039}   \\
             {1.4~~} {0.548}
            &{1.38}&{0.021} &{2.99}
            &{11.75}&{10.81} &{0.53} &{0.41}
            &{1.437} &{0.033}   \\
             {1.5~~} {0.590}
            &{1.48}&{0.019} &{2.63}
            &{11.65}&{10.82} &{0.47} &{0.36}
            &{1.581} &{0.027}   \\
             {1.6~~} {0.637}
            &{1.58}&{0.017} &{2.27}
            &{11.54}&{10.81} &{0.42} &{0.31}
            &{1.729} &{0.023}   \\
             {1.7~~} {0.693}
            &{1.69}&{0.014} &{1.95}
            &{11.39}&{10.75} &{0.37} &{0.27}
            &{1.880} &{0.019}   \\                                            {1.8~~} {0.762}
            &{1.79}&{0.012} &{1.60}
            &{11.19}&{10.64} &{0.32} &{0.23}
            &{2.035} &{0.015}   \\
             {1.9~~} {0.858}
            &{1.89}&{0.0099} &{1.29}
            &{10.91}&{10.45} &{0.27} &{0.19}
            &{2.191} &{0.011}   \\
             {2.0~~} {1.039}
            &{1.99}&{0.0070} &{0.92}
            &{10.39}&{10.04} &{0.21} &{0.15}
            &{2.337} &{0.008}   \\
\hline\hline

\end{tabular}
\end{center}
\end{table}

\begin{table}[!htb]
\begin{center}
\caption{Same with Table 2, but with the BSk21 NS EoS~\citep{chamel11,pearson12,fantina13,potekhin13}, which is based on the widely-used Skyrme nuclear energy density functional, and the parameters were fitted to reproduce with high accuracy almost all known nuclear masses, and to various physical conditions
including the nuclear matter EoS from microscopic calculations
 (see text for details).}\label{BSk21}
\begin{tabular}{cc ccc cccc cc}
\hline\hline
            {Mass~}   {Cent.}
            &\multicolumn{3}{c}{Mass }
            &\multicolumn{4}{c}{Radius}
            &\multicolumn{2}{c}{Moment of inertia} \\
            &{core}&{icrust} &{ocrust}
            &{total}&{core} &{icrust} &{ocrust}
            &{total}&{fraction} \\
\hline
             {1.0~~} {0.328}
            &{0.97}&{0.025} &{5.57}
            &{12.50}&{10.93} &{0.84} &{0.73}
            &{1.001} &{0.050}   \\
             {1.1~~} {0.346}
            &{1.08}&{0.023} &{5.08}
            &{12.54}&{11.13} &{0.76} &{0.65}
            &{1.154} &{0.043}   \\
             {1.2~~} {0.365}
            &{1.18}&{0.022} &{4.68}
            &{12.58}&{11.30} &{0.69} &{0.58}
            &{1.315} &{0.037}   \\
             {1.3~~} {0.384}
            &{1.28}&{0.021} &{4.25}
            &{12.60}&{11.44} &{0.63} &{0.53}
            &{1.482} &{0.032}   \\
             {1.4~~} {0.405}
            &{1.38}&{0.019} &{3.93}
            &{12.62}&{11.57} &{0.57} &{0.48}
            &{1.657} &{0.028}   \\
             {1.5~~} {0.428}
            &{1.48}&{0.018} &{3.57}
            &{12.62}&{11.67} &{0.52} &{0.43}
            &{1.839} &{0.025}   \\
             {1.6~~} {0.453}
            &{1.58}&{0.016} &{3.27}
            &{12.61}&{11.74} &{0.48} &{0.39}
            &{2.026} &{0.021}   \\
             {1.7~~} {0.480}
            &{1.68}&{0.015} &{2.93}
            &{12.58}&{11.79} &{0.43} &{0.35}
            &{2.220} &{0.018}   \\                                            {1.8~~} {0.511}
            &{1.79}&{0.014} &{2.63}
            &{12.52}&{11.82} &{0.39} &{0.31}
            &{2.420} &{0.016}   \\
             {1.9~~} {0.548}
            &{1.89}&{0.012} &{2.34}
            &{12.44}&{11.81} &{0.35} &{0.28}
            &{2.624} &{0.014}   \\
             {2.0~~} {0.593}
            &{1.99}&{0.011} &{2.05}
            &{12.32}&{11.76} &{0.31} &{0.25}
            &{2.833} &{0.012}   \\
\hline\hline

\end{tabular}
\end{center}
\end{table}

\begin{table}[!htb]
\begin{center}
\caption{Same with Table 3, but with the BSk20 NS EoS~\citep{chamel11,pearson12,fantina13,potekhin13} (see text for details).}\label{BSk21}
\begin{tabular}{cc ccc cccc cc}
\hline\hline
            {Mass~}   {Cent.}
            &\multicolumn{3}{c}{Mass }
            &\multicolumn{4}{c}{Radius}
            &\multicolumn{2}{c}{Moment of inertia} \\
            &{core}&{icrust} &{ocrust}
            &{total}&{core} &{icrust} &{ocrust}
            &{total}&{fraction} \\
\hline
             {1.0~~} {0.403}
            &{0.97}&{0.026} &{4.59}
            &{11.79}&{10.33} &{0.81} &{0.64}
            &{0.894} &{0.053}   \\
             {1.1~~} {0.427}
            &{1.08}&{0.024} &{4.15}
            &{11.80}&{10.50} &{0.73} &{0.57}
            &{1.029} &{0.045}   \\
             {1.2~~} {0.452}
            &{1.18}&{0.022} &{3.72}
            &{11.80}&{10.64} &{0.66} &{0.51}
            &{1.170} &{0.038}   \\
             {1.3~~} {0.480}
            &{1.28}&{0.020} &{3.37}
            &{11.79}&{10.75} &{0.59} &{0.45}
            &{1.318} &{0.033}   \\
             {1.4~~} {0.508}
            &{1.38}&{0.019} &{3.05}
            &{11.78}&{10.84} &{0.53} &{0.41}
            &{1.474} &{0.028}   \\
             {1.5~~} {0.536}
            &{1.48}&{0.017} &{2.73}
            &{11.76}&{10.92} &{0.48} &{0.36}
            &{1.638} &{0.024}   \\
             {1.6~~} {0.567}
            &{1.58}&{0.016} &{2.46}
            &{11.73}&{10.97} &{0.43} &{0.32}
            &{1.809} &{0.021}   \\
             {1.7~~} {0.602}
            &{1.69}&{0.014} &{2.18}
            &{11.67}&{10.99} &{0.39} &{0.29}
            &{1.987} &{0.018}   \\                                            {1.8~~} {0.643}
            &{1.79}&{0.013} &{1.94}
            &{11.58}&{10.98} &{0.35} &{0.26}
            &{2.170} &{0.015}   \\
             {1.9~~} {0.696}
            &{1.89}&{0.011} &{1.67}
            &{11.45}&{10.92} &{0.31} &{0.22}
            &{2.358} &{0.012}   \\
             {2.0~~} {0.764}
            &{1.99}&{0.0093} &{1.39}
            &{11.26}&{10.81} &{0.26} &{0.19}
            &{2.552} &{0.010}   \\
\hline\hline

\end{tabular}
\end{center}
\end{table}


\clearpage

\begin{thebibliography}{}

\bibitem[Akmal~et~al.(1998)]{apr}
Akmal, A., Pandharipande, V. R., \& Ravenhall, D. G., 1998, \prc, 58, 1804

 \bibitem[Andersson~et~al.(2012)]{crisis12}
 Andersson N., Glampedakis K., Ho W. C., \& Espinoza, C. M. 2012, \prl, 109, 241103

\bibitem[Antoniadis~et~al.(2013)]{2mass13}
Antoniadis, J., et al., 2013, \sci, 340, 6131

\bibitem[Baldo(1999)]{book} Baldo, M. 1999, in Nuclear Methods and the Nuclear Equation of State, ed. M. Baldo (Singapore: World Scientific), 1

 \bibitem[Baym~et~al.(1969)]{Baym69}
 Baym, G., Pethick, C., \& Pines, D. 1969, \nat, 224, 872

\bibitem[Baym et al.(1971)]{bps}
Baym, G., Pethick, C., \& Sutherland, D. 1971, \apj, 170, 299

\bibitem[Burgio et al.(2011)]{Li11y}
Burgio, G. F., Schulze, H.-J., \& Li, A. 2011, \prc, 83, 025804

 \bibitem[Caraveo et al.(2001)]{dis}
 Caraveo, P. A., De Luca, A., Mignani, R. P., \& Bignami, G. F. 2001, \apj, 561, 930

 \bibitem[Chamel(2013)]{crisis13}
 Chamel N. 2013, \prl, 110, 011101

\bibitem[Chamel et~al. (2011)]{chamel11}
Chamel, N., Fantina, A.~F., Pearson, J.~M., \& Goriely, S. 2011, \prc, 84, 062802

\bibitem[Danielewicz et~al.(2002)]{flow}
Danielewicz P., Lacey R., \& Lynch W.~G. 2002, \sci, 298, 1592

\bibitem[Dodson et~al.(2002)]{Dodson02}
Dodson R. G., McCulloch P. M., \& Lewis D. R. 2002, \apj, 564, L85

\bibitem[Demorest~et~al.(2010)]{2mass10}
Demorest, P. B., Pennucci, R., Ransom, S. M., Roberts, M. S. E., \& Hessels, J. W. T. 2010, \nat, 467, 1081

\bibitem[Douchin \& Haensel(2001)]{dh01}
Douchin, F., \& Haensel, P. 2001, \aap, 380, 151

\bibitem[Espinoza~et~al.(2011)]{Espinoza11}
Espinoza, C. M., Lyne, A. G., Stappers, B. W., \& Kramer, M. 2011, \mnras, 414, 1679

\bibitem[Fantina et~al. (2013)]{fantina13}
Fantina, A.~F., Chamel, N., Pearson, J.~M., \& Goriely, S. 2013, \aap, 559, A128

\bibitem[Feynman et al.(1949)]{fmt}
Feynman, R., Metropolis, F., \& Teller, E. 1949, \pr, 75, 1561

 \bibitem[Gugercinoglu \& Alpar(2014)]{Guger14}
 Gugercinoglu, E., \& Alpar, M. A. 2014, \apj, 788, L11

\bibitem[Hartle \& Thorne(1968)]{slow}
Hartle J.~B., \& Thorne K.~S. 1968, \apj, 153, 807

 \bibitem[Haskell~et~al.(2013)]{Haskell13}
 Haskell, B., Pizzochero, P. M., \& Seveso, S. 2013, \apj, 764, L25

 \bibitem[Hooker~et~al.(2015)]{Hooker15}
 Hooker, J., Newton, W. G., \& Li, B. A. 2015, \mnras, 449, 3559

\bibitem[Kurkela~et~al.(2010)]{Kurkela2010}
Kurkela, A., Romatschke, P., \& Vuorinen, A. 2010, \prd, 81, 105021

\bibitem[Hu et al.(2014)]{Li14y}
Hu, J. N., Li, A., Toki, H., \& Zuo, W. 2014, \prc, 89, 025802

\bibitem[Lattimer \& Lim(2013)]{lattimer13}
Lattimer, J. M., \& Lim, Y. 2013, \apj, 771, 51

\bibitem[Lattimer \& Prakash(2007)]{Lattimer2007}
Lattimer, J. M., \& Prakash, M. 2007, \pr, 442, 109

\bibitem[Li(2015)]{Li15}
Li, A. 2015, \cpl, 32, 079701

\bibitem[Li et al.(2006)]{Li06k}
Li, A., Burgio, G. F., Lombardo, U., \& Zuo, W. 2006, \prc, 74, 055801

\bibitem[Li et al.(2015)]{Li15q}
Li, A., Peng, G. X., \& Zuo, W. 2015, \prc, 91, 035803

\bibitem[Li et al.(2015a)]{Li15a}
Li, A., Shang, X. L., \& Lu, X. C. in progress

\bibitem[Li et al.(2015b)]{Li15b}
Li, A., Wang, J. B., Shao, L. J., \& Xu, R. X. 2015, \aas, 56, ...

\bibitem[Li et al.(2010)]{Li10k}
Li, A., Zhou, X. R., Burgio, G. F., \& Schulze, H.-J. 2010, \prc, 81, 025806

\bibitem[Li \& Schulze(2008)]{LS08}
Li, Z. H., \& Schulze, H. J. 2008, \prc, 78, 028801

\bibitem[Link~et~al.(1999)]{Link99}
Link, B., Epstein, R. I., \& Lattimer, J. M. 1999, \prl, 83, 3362

\bibitem[Link(2014)]{Link14}
Link, B. 2014, \apj, 789, 141

 \bibitem[Manchester et al.(2005)]{spin}
Manchester, R. N., Hobbs, G. B., Teoh, A., \& Hobbs, M. 2005, VizieR Online Data Catalog, 7245, 105

\bibitem[Negele \& Vautherin(1973)]{nv}
Negele, J. W., \& Vautherin, D. 1973, \npa, 207, 298

\bibitem[Pearson et~al. (2012)]{pearson12}
Pearson, J.~M., Chamel, N., Goriely, S., \& Ducoin, C. 2012, \prc, 85, 065803

\bibitem[Peng et al.(2008)]{Li08q}
Peng, G. X., Li, A., \& Lombardo, U. 2008, \prc, 77, 065807

\bibitem[Piekarewicz ~et~al.(2014)]{Piekarewicz14}
Piekarewicz, J., Fattoyev, F. J., \& Horowitz, C. J. 2014, \prc, 90, 015803

\bibitem[Pizzochero(2011)]{Pizzochero11}
Pizzochero, P. M. 2011, \apj, 743, L20

\bibitem[ Potekhin et~al. (2013)]{potekhin13}
Potekhin, A.~Y., Fantina, A.~F., Chamel, N., Pearson, J.~M., \& Goriely, S. 2013, \aap, 560, A48

\bibitem[ Shapiro \& Teukolsky(1983)]{tov}
Shapiro, S.~L., \& Teukolsky, S.~A. 1983, {Black holes, white dwarfs, and neutron stars: The physics of compact objects} (New York: Wiley-Interscience)

\bibitem[Sharma et al.(2015)]{bcpm}
Sharma, B. K., Centelles, M., Vinas, X., Baldo, M., \& Burgio, G. F., arXiv:1506.00375

\bibitem[Yu~et~al.(2013)]{Yu13}
Yu M., Manchester R. N., Hobbs G., Johnston S., Kaspi V. M., Keith M., Lyne A. G., Qia G. J., Ravi V., Sarkissian J. M., Shannon R., \& Xu R. X. 2013, \mnras, 429, 688

\bibitem[Zuo et al.(2004)]{Li04k}
Zuo, W., Li, A., Li, Z. H., \& Lombardo, U. 2004, \prc, 70, 055802

\clearpage
\end{thebibliography}
\end{document}